\documentstyle[subeqn,preprint,aps,epsfig]{revtex}
\begin{document}
\def\state{\langle u,u^\dagger,{\cal A}|}
\def\statetwo{|u,u^\dagger,{\cal A}\rangle\!\rangle}
\newcommand{\be}{\begin{equation}}
\newcommand{\ee}{\end{equation}}
\draft
\tighten
\title{The Schr\"odinger Representation for Fermions and a Local Expansion of the Schwinger Model}
\author{Paul Mansfield and David Nolland}
\address{Department of Mathematical Sciences,\\ Durham University,\\ 
South Road, Durham\\ DH1 3LE, U.K.\\{\tt p.r.w.mansfield@durham.ac.uk,  d.j.nolland@durham.ac.uk}}
\maketitle 
\begin{abstract}

We discuss the functional representation of fermions, and obtain exact expressions for wave-functionals of the Schwinger model. Known features of the model such as bosonization and the vacuum angle arise naturally. Contrary to expectations, the vacuum wave-functional does not simplify at large distances, but it may be reconstructed as a
large time limit of the Schr\"odinger functional, which has an expansion in
{\em local} terms.  The functional Schr\"odinger equation reduces to a set of algebraic equations for the coefficients of these terms. These ideas generalize to a numerical approach to $QCD$ in higher dimensions.

\end{abstract}

\section{Introduction}

Although the Schr\"odinger representation is a natural context for developing
non-perturbative methods in field theory, and provides useful analogs of
quantum mechanical techniques and concepts, it has not received the attention
it deserves. In particular, the existence of the Schr\"odinger equation was
not shown until the work of Symanzik \cite{1}. However there has been a
growing interest in the subject, as field theorists attempt to relate
increasingly sophisticated non-perturbative ideas to the more concrete
calculations which may be performed, for example, in lattice gauge
theory. The Schr\"odinger representation also arises naturally in the study
of field theories on space-times with boundaries \cite{2}, and has recently
been applied to investigation of the AdS/CFT correspondence\cite{3}.

Symanzik's work was restricted to perturbation theory, but in principle the
Schr\"odinger equation can be solved non-perturbatively. In  \cite{4}, it was
shown how to reduce the Schr\"odinger equation for scalar fields to an
infinite set of coupled algebraic equations. Wave-functionals (WF's) are
non-local in general, but in the presence of a mass gap they undergo a
simplification; for slowly-varying fields they have a derivative expansion, the terms of which are local expressions
involving the fields and their derivatives at a single space-time point. By
exploiting analyticity in a complex scale parameter, it was shown in \cite{4}
how this expansion could be used to reconstruct physics at all length
scales. In this paper we extend these ideas to fermionic fields. 

Since the physical spectrum of the Schwinger model has a mass gap, we
calculated its vacuum WF in the expectation that it would have a local
expansion of the form just described. However, we will find that the vacuum WF
contains non-physical massless modes which enforce gauge invariance by
screening certain large-distance configurations; because of this, the vacuum
WF cannot be expanded in local terms. This failure of the local expansion
property seems to be a general feature of gauge-invariant theories; for
example a similar result was found for Yang-Mills in \cite{Nair}. In spite of
this, however, we can still obtain
the vacuum WF from a local expansion by re-expressing it as the large-time
limit of the Schr\"odinger WF. The latter has a local expansion for {\emph
  all} field theories, even those containing massless particles, because its
time parameter acts as an inverse mass for all propagating modes. We will
verify this explicitly for the Schwinger model. 

The Schwinger model has received extensive study over
the past several decades, largely because it illustrates many features of
QCD, such as confinement and chiral symmetry breaking, in an understandable
way. It also provides a useful model for testing new techniques. 

This paper is organized as follows. In section 2 we discuss the Schr\"odinger
representation for fermions coupled to gauge fields.
Extending the discussion of the fermion representation given by Floreanini and Jackiw in \cite{5}, we argue that WF's depend on fields which are constrained by
a local chiral-type projection implemented by operators $Q_\pm$. We identify all
possible choices of $Q_\pm$ in arbitrary dimension, arguing that they are
parametrized by a single complex number. We find a path-integral expression
for the vacuum WF of a gauge theory coupled to fermions. In section 3 we explicitly obtain the vacuum WF of the
Schwinger model in two ways, by solving the path-integral, and by solving the
Schr\"odinger equation. The path-integral solution provides a particularly
straightforward derivation of this result. We discuss various physical issues such as the vacuum
angle, 
bosonization, confinement, and vacuum condensates. In section 4 we obtain the
Schr\"odinger functional,  verify that it has an expansion in local terms, and show that
its large time limit reproduces the result of the preceding section. In
section 5 we present our conclusions and discuss the significance of these
results in the context of a proposed numerical approach to higher dimensional $QCD$. 
   
\section{Representing Fermion Fields}

Wave-functionals for fermions coupled to a gauge field may be represented as overlaps with a field state. We work in the Weyl gauge $A_0=0$, so that the arguments of WF's are fields (Grassman-valued for fermions) defined on a  surface of constant time, and WF's are invariant under time-independent gauge transformations of these fields. So for the vacuum WF, for example, we have
\be \Psi[u,u^\dagger,{\cal A}]=\state 0\rangle,\label{over} \ee
where we have suppressed vector, spinor, colour and flavour indices. We will
find that $u$ and $u^\dagger$ are actually constrained fields, subject to a
local chiral-type projection. 
The field state is chosen as follows: it is diagonal in the gauge field
\be\state\hat A={\cal A}\state,\label{vec}\ee
and the conjugate electric field is represented by functional differentiation
\be\state\hat E=-i{\delta\over\delta{\cal A}}\state.\ee

For fermions we use the Florenanini-Jackiw representation \cite{5}

\begin{eqnarray}
\langle\!\state\hat{\psi }  & = & \frac{1}{\sqrt{2}}(u+\frac{\delta }{\delta u^{\dagger }})\langle\!\state\nonumber \\
\langle\!\state\hat{\psi }^{\dagger }  & = & \frac{1}{\sqrt{2}}(u^{\dagger }+\frac{\delta }{\delta u})\langle\!\state.\label{rep} 
\end{eqnarray}

The notation of (\ref{rep}) indicates that $u$ and $u^\dagger$ are
temporarily treated as
unconstrained fields, eg. for the purpose of taking functional derivatives. It is
obvious that (\ref{rep}) is reducible, but it is important to note that its reducibility may be removed by imposing
the constraints  \( Q_{-}u=u^{\dagger }Q_{+}=0 \), for some arbitrary
projection operators \( Q_{\pm }=\frac{1}{2}(1\pm Q) \), \( Q^{2}=1 \). In
the presence of these constraints, the representation (\ref{rep}) corresponds
to diagonalizing $Q_+\psi$ and $\psi^\dagger Q_-$:

\begin{eqnarray}
\state Q_{+}\hat{\psi }  & = & \sqrt{2}Q_{+}u\state \nonumber \\
\state\hat{\psi }^{\dagger }Q_{-}  & = & \sqrt{2}u^{\dagger }Q_{-}\state ,\label{frep} 
\end{eqnarray}

and representing the other projections by functional differentiation

\begin{eqnarray}
\state Q_{-}\hat{\psi }  & = & \frac{1}{\sqrt{2}}Q_{-}\frac{\delta }{\delta u^{\dagger }}\state \nonumber \\
\state\hat{\psi }^{\dagger }Q_{+}  & = & \frac{1}{\sqrt{2}}\frac{\delta }{\delta u}Q_{+}\state .\label{frep2} 
\end{eqnarray}

We
will sometimes use (\ref{rep}) for
 convenience, but this is equivalent to using (\ref{frep}) and (\ref{frep2}). So physical WF's
will always depend only on the constrained fields, and in our final results
the constraints will always be assumed.

We can make the field dependence of the field state more explicit by
writing\footnote{In this section traces are functional, ie. over spatial
  indices as well as colour and flavour, if present.
}

\be
\label{sta}
\state=e^{-{\rm Tr}(u^\dagger Qu)}\langle\!\state = \langle Q|\exp {\mathrm Tr}[\sqrt{2}(u^{\dagger }Q_-\hat{\psi }-\hat{\psi }^{\dagger }Q_+u)+i{\cal A}\cdot \hat{E}] ,
\ee
Note the relative factor of $e^{-{\rm Tr}(u^\dagger Qu)}$ between $\state$
and $\langle\!\state$, which disappears when the constraints are imposed. \(
\langle Q|  \) is defined by 
\begin{equation}
\label{c1}
\langle Q|\hat{A} =\langle Q|Q_{+}\hat{\psi } =\langle Q|\hat{\psi }^{\dagger }Q_{-} =0.
\end{equation}

Similarly, defining $|u,u^\dagger,{\cal A}\rangle=\exp{\rm
  Tr}[\sqrt2(u^\dagger\hat\psi-\hat\psi^\dagger u)+i{\cal A}\cdot\hat
  E]|Q\rangle$ with $|Q\rangle$ satisfying $\hat A|Q\rangle=Q_-\hat\psi|Q\rangle=\hat\psi^\dagger
  Q_+|Q\rangle=0$, we have
\begin{eqnarray}
\hat{\psi }\statetwo  & = & \frac{1}{\sqrt{2}}(u-\frac{\delta }{\delta u^{\dagger }})\statetwo\nonumber \\
\hat{\psi }^{\dagger }\statetwo  & = & \frac{1}{\sqrt{2}}(u^{\dagger }-\frac{\delta }{\delta u})\statetwo.\label{rep2} 
\end{eqnarray} 

Next we wish to define an inner-product. The definitions (\ref{sta})
and (\ref{c1}) give rise to the following equations:
\begin{eqnarray}
0 & = & (\hat{\psi }^{\dagger }-\sqrt{2}u^{\dagger })Q_+\langle\!\langle u,u^{\dagger },{\cal A}|{v},{v}^{\dagger },\tilde{\cal A}\rangle\!\rangle \nonumber \\
 & = & Q_{+}(\hat{\psi }+\sqrt{2}{v})\langle\!\langle u,u^{\dagger },{\cal A}|{v},{v}^{\dagger },\tilde{\cal A}\rangle\!\rangle \nonumber \\
 & = & (\hat{\psi }^{\dagger }+\sqrt{2}{v}^{\dagger })Q_-\langle\!\langle u,u^{\dagger },{\cal A}|{v},{v}^{\dagger },\tilde{\cal A}\rangle\!\rangle \nonumber \\
 & = & Q_{-}(\hat{\psi }-\sqrt{2}u)\langle\!\langle u,u^{\dagger },{\cal A}|{v},{v}^{\dagger },\tilde{\cal A}\rangle\!\rangle ,\label{e4} 
\end{eqnarray}
where we have used the canonical commutation relations. The field operators
in (\ref{e4}) may be represented by either pair of fields, and the equations
solved to give

\begin{equation}
\label{bound}
\langle\!\langle u,u^{\dagger },{\cal A}|{v},{v}^{\dagger },{\cal A}\rangle\!\rangle =\exp {\mathrm{Tr}}(u^{\dagger }Qu-u^{\dagger }2Q_{-}{v}+{v}^{\dagger }2Q_{+}u+{v}^{\dagger }Q{v})\delta({\cal A}-\tilde{\cal A}).
\end{equation}

Hence
\be
\state v,v^\dagger,\tilde{\cal A}\rangle = \exp{\rm Tr}(2v^\dagger
u-2u^\dagger v)\delta({\cal A}-\tilde{\cal A}),
\ee

which differs slightly from the inner-product given in \cite{5} and
elsewhere. To calculate inner-products of WF's we use the partition of unity
\be 
\label{unity}
1=\int D{\cal A}DuDu^\dagger DvDv^\dagger e^{{\rm Tr}(2v^\dagger
u-2u^\dagger v)}|v,v^\dagger,{\cal A}\rangle\state.
\ee

Now for many purposes it is convenient to require physical WF's to be
gauge invariant, so we wish to choose \( Q \) in such a way that this is
made possible.
In general, we could consider \( Q \) to have some gauge field dependence.
However, any field dependence, or non-locality of
$Q$ will cause the representation to transform non-trivially under
time-independent gauge
transformations (this is easily seen from (\ref{sta}) and (\ref{c1})). The
only gauge invariant representations are when we take \( Q \) to be a local,
field independent operator.  

Several authors, inspired by the similarity of \( Q \) with a choice of Dirac
sea, have taken \( Q_{\pm } \) to be (non-local) projections \( P_{\pm } \) onto +ve/-ve energy eigenstates.
The resulting WF's are invariant under time-independent gauge transformations
of the fields, but in general they do not satisfy Gauss' law \cite{5,6}. However, if we
take \( Q \) to satisfy the conditions above, Gauss' law is always
automatically satisfied.

We can obtain further restrictions on the choice of \( Q \) by requiring that
the overlap (1) is well-defined and non-vanishing; since \( |u,u^{\dagger },{\cal A}\rangle  \)
is not a physical state this is not guaranteed to be the
case. Consider the vacuum. With $P_\pm$ defined as above, it satisfies
\be\hat\psi^\dagger P_+|0\rangle=P_-\hat\psi|0\rangle=0.\label{c2}\ee
The conditions (\ref{c1}) and (\ref{c2}) and the representation (\ref{rep}) lead to the following equations:
\begin{eqnarray}& &P_-\Bigl(u+{\delta\over\delta u^\dagger}\Bigr)\langle\!\state0\rangle=0\nonumber\\& &\Bigl(u^\dagger+{\delta\over\delta u}\Bigr)P_+\langle\!\state0\rangle=0\nonumber\\& &Q_+\Bigl(u-{\delta\over\delta u^\dagger}\Bigr)\langle\!\state0\rangle=0\nonumber\\& &\Bigl(u^\dagger-{\delta\over\delta u}\Bigr)Q_-\langle\!\state0\rangle=0.\label{eqn}
\end{eqnarray}
These have the solution 
\be\langle\!\state0\rangle=\langle Q|Be^{u^\dagger Mu}\exp{\rm Tr}\Bigl (i{\cal A}\cdot\hat E\Bigr)|0\rangle,\label{sol}\ee
where $M=A^{-1}C=C^{-1}A$, and we define $A=\{P_-,Q_+\}$, $C=[P_-,Q_+]$. The
constant of integration $B$ corresponds to the determinant of the Dirac operator, and solving (\ref{eqn}) is equivalent to performing the fermion integration, since (\ref{sol}) no longer involves fermion field operators. We assume that $A^{-1}$ exists; we can always choose $Q_\pm$ so that this is the case.

We can reproduce this result in a path integral representation. The vacuum is
represented as $\lim_{t\to\infty}e^{-\hat Ht}|S\rangle$, where $S$ is any
physical state not orthogonal to the vacuum. This leads, via the usual correspondence of matrix elements with path integrals, to the expression
\begin{eqnarray}\Psi[u,u^\dagger,{\cal A}]=& &\int DA_iD\psi^\dagger D\psi\exp\biggl\{-S_B-S_F\nonumber\\& &+{\rm Tr}\Bigl[\sqrt2(u^\dagger Q_-\psi-\psi^\dagger Q_+u)+u^\dagger Qu+i{\cal A}\cdot\dot A\Bigr]\biggr\}.\label{path}\end{eqnarray}
Here $S_B$ and $S_F$ are the bosonic and fermionic parts of the Euclidean
action, and the integral is evaluated with the following boundary conditions implied by (\ref{c1}):
\be A_i|_{t=0}=Q_+\psi|_{t=0}=\psi^\dagger Q_-|_{t=0}=0.\ee  

The easiest way to impose the boundary conditions is by point splitting the
equal time propagators which arise in (\ref{path}). We choose for
$\psi^\dagger Q_+$ and $Q_-\psi$ to be propagated forward in time, and
$\psi^\dagger Q_-$, $Q_+\psi$ backwards (the direction of propagation of the
gauge field is arbitrary). This is effected by adding boundary terms 
\be
(\psi^\dagger_{-\epsilon}Q_+\psi)+i\rm{Tr}(A^{-\epsilon}\cdot\dot A)
\label{bterms}
\ee
to the action, where $\psi^\dagger_{-\epsilon}(\underline
x,t)=\psi^\dagger(\underline x,t-\epsilon)$, etc. and we let $\epsilon\rightarrow 0$. After adding these terms, we are free to choose any boundary conditions we like, or to integrate freely over the boundary values; the result is equivalent to evaluating (\ref{path}) with the boundary conditions implied by (\ref{c1}).

The fermion integration is now easily done by standard methods; we obtain the path integral version of (\ref{sol})
\be\Psi[u,u^\dagger,{\cal A}]=\int DA_i\det D\ \exp\biggl\{-S_B+{\rm Tr}\Bigl[u^\dagger A^{-1}Cu+i({\cal A}-A)\cdot\dot A\Bigr]\biggr\}.\label{soln}\ee
$D$ is the Dirac operator. We can also write down the vacuum functional for free fermions or fermions in a classical background field:
\be\Psi[u,u^\dagger]=\exp{\rm Tr}\Bigl[u^\dagger A^{-1}Cu\Bigr].\ee

We will now investigate the possible choices for $Q$. The solution we found
above depended on the existence of the operator $A^{-1}C$; for local $Q$
if this does not exist, (\ref{eqn}) has no non-singular solution. We can
write $A^{-1}C=(Q+P)^{-1}P(Q+P)$, so we need $Q+P$ to be invertible for all
momenta. All $Q$ satisfying this condition are given in Appendix E of
\cite{1}; they take the form 

\be 
\label{posq}
Q=a\gamma^0+ib\gamma^0\gamma^5+c\gamma^5\label{genq}
\ee
where $a^2+b^2+c^2=1$, with the additional conditions
\be
(1-b^2)^{-1}(-a\pm ibc)\rlap{\ /}\in[1,\infty),
\ee
for $m>0$, and $b^2\ne1$ for $m=0$.

We can rewrite (\ref{posq}) as 
\be Q_+=(a\gamma^0+b\gamma^1){1\over2}(1+a^\prime\gamma^0+b^\prime\gamma^1),\ee 
\be
Q_-={1\over2}(1-a^\prime\gamma^0-b^\prime\gamma^1)(a\gamma^0+b\gamma^1),\ee
where $a^\prime={a+ibc\over a^2+b^2}$ and $b^\prime={b-iac\over a^2+b^2}$;
from (\ref{frep}) and (\ref{frep2}) we see that this is equivalent to choosing
$Q_\pm={1\over2}(1\pm a^\prime\gamma^0\pm b^\prime\gamma^1)$. Since
$a^{\prime2}+b^{\prime2}=1$, this means that we can set $c=0$ in (\ref{genq})
without loss of generality. Thus the possible choices for $Q$ are
parametrized by the single, arbitrary complex number $z=a-ib$ (which could in
principle be taken to have local variations in space). 

Specializing to two dimensions we can take $\gamma^0=\sigma^1$, $\gamma^1=\sigma^2$, $\gamma^5=\sigma^3$, in which case
\be Q_\pm={1\over2}\pmatrix{1&\pm z\cr\pm z^{-1}&1}.\ee

In two dimensions, the general form of a physical WF is
\begin{equation}
\label{wf}
\Psi =\state\Psi_f\rangle=\sum _{a=0}^{\infty }\frac{1}{a!}\prod _{n=1}^{a}\int dx_{n}dy_{n}u^{\dagger }(x_{n})\gamma^5u(y_{n})e^{ie\int ^{x_n}_{y_n}{\cal A}}f^{a}(x_{1},y_{1},\ldots ,x_{a},y_{a}),
\end{equation}
where the Wilson lines $e^{ie\int ^{x_n}_{y_n}{\cal A}}$ guarantee gauge
invariance and $f^a$ are arbitrary functions. It is useful to note that in
the presence of the constraints on $u$ and $u^\dagger$, two dimensional gamma
matrices have only one effective degree of freedom; ie. in (\ref{wf}) we
could have written $\gamma^0$ or $\gamma^1$ in place of $\gamma^5$.

Using this and (\ref{unity}) we can calculate (dropping a divergent constant)
\be
\langle\Psi_g|\Psi_f\rangle=g^0f^0+\sum_{a=1}^\infty{1\over a!}\int
d^axd^ay\epsilon_{i_1\ldots
  i_a}g^a(x_1,y_{i_1},\ldots,x_a,y_{i_a})^*f^{a}(x_{1},y_{1},\ldots
,x_{a},y_{a}).
\ee

\section{The Schwinger Model Vacuum}
\subsection{Path Integral Solution}

In (1+1) dimensions gauge theories are sufficiently simple to admit analytical treatment, and provide a great deal of information about the general structure of WF's. Consider the Schwinger model, with Euclidean action
\be
\label{sact}
\int d^2x(\bar\psi\gamma\cdot(\partial+ieA)\psi+{i\over4}F^{\mu\nu}F_{\mu\nu}).
\ee
We take gamma matrices to be Euclidean,
${\gamma^\mu,\gamma^\nu}=\delta^{\mu\nu}$. For massless fermions in (1+1) dimensions we can obtain the fermion determinant by integrating the anomaly \cite{7}; with a U(1) gauge group the result is
\be
\label{fdet}
\det D=\exp\Bigg\{-{e^2\over2\pi}(\partial_\mu\phi)^2\Bigg\},
\ee
where $A_\mu=\partial_\mu\eta+\epsilon_{\mu\nu}\partial_\nu\phi$. $P_\pm$ are obtained as equal time limits of the Dirac propagator 
\be P_\pm=\lim_{\delta t\rightarrow 0^\pm}D^{-1},\ee 
and
\be D=e^{ie(\eta+\gamma^5\phi)}\gamma^0\gamma.\partial e^{-ie(\eta+\gamma^5\phi)}\ee 
At $t=0$ we want $A_1={\cal A}$, which implies that $\eta=\int^x({\cal
  A}+\dot\phi)$. For $\phi$ we choose the boundary condition $\phi|_{t=0}=0$,
corresponding to the vanishing of the vacuum angle, as we will explain
later. Thus we have 
\be P_\pm={1\over2}\Biggl(\delta(x-y)\pm{\gamma^5\over\pi}{\cal P}{1\over x-y}\Biggr)\exp\Bigl\{ie\int^x_y({\cal A}+\dot\phi)\Bigr\}.\ee
Here ${\cal P}$ denotes the principal part. The operator $A^{-1}C$ is easily found; (\ref{soln}) becomes
\begin{eqnarray}\Psi[u,u^\dagger,{\cal A}]=& &\int D\phi\exp\Biggl\{-{1\over2}\phi(\partial^4+M^2\partial^2)\phi\nonumber\\& &+{2\over\pi}\int dxdy\biggl[u^\dagger(x)\gamma^5u(y){\cal P}{1\over x-y}\exp\Bigl\{ie\int^x_y({\cal A}+\dot\phi)\Bigr\}\biggr]\Biggr\},\label{int}\end{eqnarray}
where $M^2={e^2\over\pi}$. We can integrate this to give
\begin{eqnarray}\Psi[u,u^\dagger,{\cal A}]=& &\sum^\infty_{a=0}{1\over a!}\prod^a_{n=0}\Biggl[{2\over\pi}\int dx_ndy_nu^\dagger(x_n)\gamma^5u(y_n){\cal P}{1\over x_n-y_n}e^{ie\int^x_y{\cal A}}\nonumber\\& &\times\exp\biggl\{\sum^a_{i,j=1}\Phi(x_i-y_i)-\sum^a_{j>i=1}\Bigl[\Phi(x_i-x_j)+\Phi(y_i-y_j)\Bigr]\biggr\}\Biggr],\label{vac}\end{eqnarray}where
\be\Phi(x)=\int{dp\over2\pi}\Biggl({1\over|p|}-{\sqrt{p^2+M^2}\over p^2}\Biggr)(1-\cos px).\label{potential}\ee
It may be explicitly verified that (\ref{vac}) satisfies both Gauss' law and the Schr\"odinger equation. Note that this expression is both UV and IR convergent; apart from a subtraction of zero-point energy, no renormalization is necessary in this model.

\subsection{The Vacuum Angle}

The WF (\ref{vac}) is strictly gauge-invariant, so it corresponds to
the vacuum with vanishing vacuum angle. However, a non-zero vacuum angle may be
obtained by inserting a term 

\be
\label{vterm}
{ie\theta\over2\pi}\int
dx\epsilon^{\mu\nu}F_{\mu\nu}
\ee
 into the path-integral expression
(\ref{path}). This term is proportional to the ``instanton number'', which is
{\em not\/} quantized, as a result of the boundary conditions. We have
$\epsilon^{\mu\nu}F_{\mu\nu}=-2\partial^2\phi$, but on the other hand ${\cal
  A}=-\int_{-\infty}^0dt\partial^2\phi$, so that \be
\label{fa}
\int
d^2x\epsilon^{\mu\nu}F_{\mu\nu}=2\int dx{\cal A}.
\ee
Hence
\be
\Psi_\theta[u,u^\dagger,{\cal A}]=\Psi[u,u^\dagger,{\cal
  A}]e^{{ie\theta\over\pi}\int dx{\cal A}}
\ee
so that under a large gauge transformation $\int dx{\cal A}\rightarrow\int dx{\cal
  A}+{2\pi n\over e}$, we have $\Psi_\theta\rightarrow\Psi_\theta
e^{2in\theta}$. This phase is twice what we might expect; because of the relation
(\ref{fa}) large gauge transformations change the instanton number by
multiples of two. To change the instanton number by {\em one\/} unit, we must perform a
large gauge transformation of winding-number one half $\int dx{\cal A}\rightarrow\int dx{\cal
  A}+{\pi\over e}$. These ``half-integer'' transformations were first discussed
in \cite{8}, but should not be thought of as contributing to the
vacuum degeneracy \cite{9}. The periodicity of
$\theta$ is explained by the fact that it is energetically favourable to
create an electron-positron pair when $\theta>2\pi$ \cite{10}.

Because of the chiral anomaly, there is another representation for the
theta vacuum;
by performing a chiral rotation
\be
\label{crot}
\psi\rightarrow e^{i\theta\gamma^5/2}\psi,
\qquad\psi^\dagger\rightarrow\psi^\dagger e^{-i\theta\gamma^5/2},
\ee
in (\ref{path}) the term (\ref{vterm}) is cancelled. This returns us to the original
expression, but with $Q$ replaced by $Q^\prime=Qe^{i\theta\gamma^5}$ and $u$,
$u^\dagger$ chirally rotated, ie.
\be
\Psi_\theta[u,u^\dagger,{\cal A}]=\Psi[e^{-i\theta\gamma^5/2}u,u^\dagger e^{i\theta\gamma^5/2},{\cal
  A}].
\ee
This implies that 
\be
\label{vacrep}
\langle\Psi^1_\theta|\hat
O(\hat\psi,\hat\psi^\dagger)|\Psi^2_\theta\rangle=\langle\Psi^1|\hat
O( e^{i\theta\gamma^5/2}\hat\psi,\hat\psi^\dagger
e^{-i\theta\gamma^5/2})|\Psi^2\rangle,
\ee
a fact which is useful for calculating expectation values in the
$\theta$-vacuum. 

All of these results are unaffected by giving the fermions a non-zero mass,
except that the mass term is also altered by the chiral rotation
(\ref{crot}); $m\rightarrow me^{i\theta\gamma^5}$. This indicates that the
vacuum angle has a true physical significance in this case, whereas for
massless fermions it can be altered by a suitable redefinition of the fields.

\subsection{Solution from the Schr\"odinger Equation }

A solution for the vacuum WF of the Schwinger model (on a circle) was given in \cite{11},
where it was found by solving the Schr\"odinger equation. We now wish to show
how a similar calculation can reproduce the result of the last section. Define
\begin{equation}
\label{jay}
\hat{J}_{M}(x,y)=\hat{\psi }^{\dagger }(x)M\hat{\psi }(y)e^{ie\int ^{x}_{y}\hat{A}}.
\end{equation}
Gauss' law implies that \( \partial _{x}\hat{E}(x)\sim e\hat{j}_{0}(x) \) on
physical states, where \( \hat{j}_{0}=\hat{\psi }^{\dagger }\hat{\psi } \).
By Fourier transforming, this in turn gives us
\begin{equation}
\label{gauss}
\int dx\hat{E}^{2}(x)\sim \int
dxdy\hat{j}_{0}(x)\hat{j}_{0}(y)\bar{V}(x-y)+\hat F^2,
\end{equation}
where $\hat F=\int dx \hat E$ and
\begin{equation}
\label{pot}
\bar{V}(x)=\frac{e^{2}}{2\pi }\int \frac{dp}{p^{2}}e^{ipx}.
\end{equation}

Thus the Hamiltonian acting on physical states can be written as
\begin{equation}
\label{ham}
\hat{H}=\int dxdy\hat{j}_{0}(x)\hat{j}_{0}(y)\bar{V}(x-y)-i\int dx\lim
_{y\rightarrow x}\partial _{y}\hat{J}_{\gamma ^{5}}(x,y)+\hat F^2.
\end{equation}
Note that the fermion interaction is completely determined by Gauss' law. The
$\hat F$ term is represented by $\partial\over\partial(\int dx{\cal A})$, and
may be diagonalized by inserting a factor of $e^{{ie\theta\over\pi}\int
  dx{\cal A}}$ into the WF's, ie. it gives the vacuum angle. For the moment we
  will take this to vanish, so that this term may be ignored.

Gauge invariant WF's have the form (\ref{wf}). If the distributions \( f^{a} \) are taken to be
translation invariant, then (\ref{wf}) is simultaneously a zero eigenstate
of momentum. Inspired
by the path integral calculation of the last section we make the Ansatz
\begin{eqnarray}
\label{ans}
& &f^{a}(x_{1},y_{1},\ldots ,x_{a},y_{a})\nonumber\\& &\qquad=2\prod _{n=1}^{a}\Delta (x_{n}-y_{n})\exp \left\{ \sum _{i,j=1}^{a}\Phi (x_{i}-y_{i})-\sum _{j>i=1}^{a}\left[ \Phi (x_{i}-x_{j})+\Phi (y_{i}-y_{j})\right] \right\} ,
\end{eqnarray}
where \( \Delta (x)=\frac{i}{\pi }{\cal P}\frac{1}{x-y} \).  The equal-time limits of the
free-fermion propagator are  \({1\over
  2}(1\pm\gamma^5\Delta)\), and \( \Phi  \) is the two-point function
of the effective theory which results from integrating out the
  fermions
.
In a more general theory we would include higher $n$-point functions in (\ref{ans}),
but there are no higher $n$-point functions in this case, as the effective theory
is non-interacting. We take \( \Phi \) symmetric and \( \Phi (0)=0 \),
so we can write
\begin{equation}
\label{defc}
\Phi (x)=\int dpC(p)(1-\cos(px)),
\end{equation}
 and we have only to determine \( C(p) \). Let us write
\begin{equation}
\label{tildef}
\hat{H}\Psi =\sum _{a=0}^{\infty }\frac{1}{a!}\prod _{n=1}^{a}\int dx_{n}dy_{n}u^{\dagger }(x_{n})\gamma ^{5}u(y_{n})e^{ie\int ^{x_{n}}_{y_{n}}\cal A}\tilde{f}^{a}(x_{1},y_{1},\ldots ,x_{a},y_{a}).
\end{equation}
We will interpret sums and products from 1 to 0 as 0 and 1 respectively. An explicit computation, as in \cite{11}, gives

\begin{eqnarray}
\tilde{f}^{a}(x_{1},y_{1},\ldots ,x_{a},y_{a}) & = & V(x_{1},y_{1},\ldots ,x_{a},y_{a})f^{a}(x_{1},y_{1},\ldots ,x_{a},y_{a})\nonumber \\
 &  & -i\int dx\lim _{y\rightarrow x}\partial _{y}f^{a+1}(x_{1},y_{1},\ldots ,x_{a},y_{a},y,x)\nonumber \\
 &  & +i\int dx\lim _{y\rightarrow x}\partial _{y}\sum _{b=1}^{a}f^{a+1}(x_{1},y_{1},\ldots ,x_{b},x,y,y_{b},\ldots ,x_{a},y_{a})\nonumber \\
 &  & -i\sum _{b=1}^{a}f^{a-1}(x_{1},y_{1},\ldots ,\rlap /x_{b},\rlap /y_{b},\ldots ,x_{a},y_{a})\delta ^{\prime }(x_{b}-y_{b}),\label{sch} 
\end{eqnarray}
where $\rlap /x$ denotes the omission of that argument, and we define
\begin{equation}
\label{pot2}
V(x_{1},y_{1},\ldots ,x_{a},y_{a})=\sum _{i,j=1}^{a}V(x_{i}-y_{i})-\sum _{j>i=1}^{a}\left[ V(x_{i}-x_{j})+V(y_{i}-y_{j})\right] ,
\end{equation}
and \( V(x)=2(\bar{V}(0)-\bar{V}(x)) \).

We wish to verify that \( \Psi  \) solves the Schr\"odinger equation
\begin{equation}
\label{schr}
\hat{H}\Psi =E_{0}\Psi .
\end{equation}
For \( a=0 \) (\ref{schr}) reduces to
\begin{equation}
\label{energy}
E_{0}=-i\int dx\lim _{y\rightarrow x}\left\{ \partial _{y}(\Delta (y-x)e^{\Phi (y-x)})\right\} ,
\end{equation}
and gives us the divergent vacuum energy. For \( a=1 \) we find
\begin{eqnarray}
E_{0}\Delta (x_{1}-y_{1}) & = & V(x_{1}-y_{1})\Delta (x_{1}-y_{1})\nonumber \\
 &  & +i\int dx\Delta (x_{1}-x)\Delta (x-y_{1})(\Phi ^{\prime }(x-y_{1})-\Phi ^{\prime }(x-x_{1}))\nonumber \\
 &  & -i\Delta (x_{1}-y_{1})\int dx\lim _{y\rightarrow x}\partial _{y}\left( \Delta (y-x)e^{g(x,y,x_{1},y_{1})}\right) ,\label{a1} 
\end{eqnarray}
where \( g(x,y,x_{1},y_{1})=\Phi (y-x)+\Phi (x_{1}-x)+\Phi (y-y_{1})-\Phi (y-x_{1})-\Phi (x-y_{1}) \), and we have used $\int dx\Delta(x_1-x)\Delta(x-x_2)=\delta(x_1-x_2)$.
Now by Taylor expanding a test function \( f(x) \) it is clear that
\begin{eqnarray}
\label{lim}
\lim _{x\rightarrow 0}\Delta (x)f(x)&=&\frac{i}{\pi }f^{\prime }(0)+\ldots\nonumber\\
\lim _{x\rightarrow 0}\Delta ^{\prime }(x)f(x)&=&\frac{-i}{2\pi }f^{\prime \prime }(0)+\ldots
\end{eqnarray}
The ellipses represent potentially divergent terms which contribute to the
vacuum energy but cancel in (\ref{a1}). Hence we have
\be
-i\int dx\lim _{y\rightarrow x}\partial _{y}\left( \Delta
  (y-x)e^{g(x,y,x_{1},y_{1})}\right)   =  E_{0}+\frac{1}{2\pi }\int dx(\Phi
^{\prime }(x-y_{1})-\Phi ^{\prime }(x-x_{1}))^{2},
\ee
and
\be
 \frac{1}{2\pi }\int dx(\Phi
^{\prime }(x-y_{1})-\Phi ^{\prime }(x-x_{1}))^{2}=\int dpp^{2}C(p)^{2}(1-\cos(p(x_1-y_1))).\label{iden} 
\ee
Now the identity
\begin{eqnarray}
\label{iden2}
\int dx\Delta(x_1-x)\Delta(x-y_1)f(x)&=&\delta(x_1-y_1)f(x_1)\nonumber\\
& &+\Delta(x_1-y_1)\int{dp\over2\pi}
\epsilon(p)\tilde f(p)(e^{-ipx_1}-e^{-ipy_1}) 
\end{eqnarray}
corresponds in momentum space to the identity
$\epsilon(p)\epsilon(p+q)=1+\epsilon(p)\epsilon(q)-\epsilon(q)\epsilon(p+q)$\\
($\epsilon$ is the step function). This implies that
 
\begin{eqnarray}
&&i\int dx\Delta (x_{1}-x)\Delta (x-y_{1})(\Phi ^{\prime }(x-y_{1})-\Phi ^{\prime }(x-x_{1}))\nonumber\\&&=2\Delta (x_{1}-y_{1})\int dp|p|C(p)(1-\cos(p(x_1-y_1))).
\end{eqnarray}

Thus (\ref{a1}) reduces to the quadratic equation
\begin{equation}
\label{quad}
\frac{e^{2}}{\pi p^{2}}+2|p|C(p)-p^{2}C(p)^{2}=0,
\end{equation}
and to get a normalizable WF we must take the appropriate root
\begin{equation}
\label{csol}
C(p)=\frac{|p|-\sqrt{p^{2}+M^{2}}}{p^{2}}.
\end{equation}
This reproduces the result
(\ref{vac}). It is straightforward to show that (\ref{schr}) is satisfied
by this solution for all \( a>1 \); using the identities (\ref{iden}) and
(\ref{iden2}) the $a=n$ equation may be reduced to multiple copies of the $a=1$
equation.

\subsection{Bosonization}

The confining nature of the theory is easily seen from (\ref{vac}); since the
potential (\ref{pot}) goes like $|x|$ at large distances, configurations in
which a field source goes off to infinity without an accompanying anti-source, are exponentially damped. Infrared slavery and asymptotic freedom are similarly seen by scaling the momentum in (\ref{pot}); the effect of such scaling is to make the coupling constant large for small momenta and vice-versa.

Bosonization may be understood by transfering the sources to currents instead
of fields. By a gauge transformation, we can always set \( {\cal A}=0 \).
Now consider the following object:
\begin{equation}
\label{current}
\Psi [\xi ]=\int Du^{\dagger }Due^{\int d^2xu^{\dagger }\gamma^5u}\Psi [u,{\sqrt\pi\over2}\xi u^{\dagger },0].
\end{equation}
Inserting the path-integral expression (\ref{int}) and
performing the \( u,u^{\dagger } \) integrations gives (with boundary conditions
as before)
\begin{equation}
\label{couple}
\Psi [\xi ]=\int D\phi D\psi ^{\dagger }D\psi e^{-S_{B}-S_{F}+\sqrt\pi\mathrm{Tr}(\xi \psi^\dagger\gamma^5\psi )}.
\end{equation}

This is a functional whose argument couples to a local fermion
current. Using (\ref{fdet}) the fermion integration yields
\begin{eqnarray}
\Psi[\xi]&=&\int D\phi e^{-S_B-{1\over2\pi}\int
  d^2x(\partial\varphi)^2}\nonumber\\
&=&\int D\phi e^{-{1\over2}\int d^2x(\partial^2\phi)^2-{1\over 2\pi}\int d^2x\bigl(e^2\phi\partial^2\phi-2e\tilde\phi\phi+\tilde\phi{1\over\partial^2}\tilde\phi\bigr)} ,
\end{eqnarray}
where \(\varphi=e\phi-{1\over\partial^2}\tilde\phi \) and
\(\tilde\phi=\epsilon_{\lambda\mu}\partial_\lambda J_\mu\)
with $J_0=0$, $J_1=i\sqrt\pi\xi\delta(t)$. Hence
\(\tilde\phi=i\sqrt\pi{\partial\over\partial
  t}\bigl(\xi\delta(t)\bigr)\) and performing
the remaining integration, we find
\begin{equation}
\label{boson}
\Psi [\xi ]=e^{-\frac{1}{2}\int dx\xi \sqrt{-\partial_x ^{2}+M^{2}}\xi },
\end{equation}
which is the vacuum WF of a free boson field of mass $M$.

From this we can easily reproduce the standard correspondence between bosonic and
fermionic operators. (\ref{couple}) implies that
\begin{eqnarray}
\label{bos}
{1\over\sqrt\pi}{\delta\over\delta\xi}\Psi [\xi ]&=&\int D\phi D\psi ^{\dagger }D\psi
\psi^\dagger\gamma^5\psi e^{-S_{B}-S_{F}+\sqrt\pi\mathrm{Tr}(\xi \psi^\dagger\gamma^5\psi )}\nonumber\\&=&\int Du^{\dagger }Due^{\int d^2xu^{\dagger }\gamma^5u}\hat\psi^\dagger\gamma^5\hat\psi\Psi [u,{\sqrt\pi\over2}\xi u^{\dagger },0].
\end{eqnarray}
The fermion
operators are represented as in (\ref{frep}) and (\ref{frep2}). Now define 
\be
\Psi [\xi,\chi]=\int D\phi D\psi ^{\dagger }D\psi
e^{-S_{B}-S_{F}+\sqrt\pi\mathrm{Tr}(\xi \psi^\dagger\gamma^5\psi
  +\chi\psi^\dagger\psi)}.
\ee
Performing the fermion integration we find
\(J_1=i\sqrt\pi\xi\delta(t)\), as before, and
\(J_0=\sqrt\pi\chi\delta(t)\), so that
\(\tilde\phi=i\sqrt\pi{\partial\over\partial
  t}(\xi\delta(t))+\sqrt\pi{\partial\over\partial t}(\chi\delta(t))\).
Thus
\be
\Psi [\xi,\chi]=e^{\int
  dx\bigl(-{1\over2}\xi\sqrt{-\partial^2+M^2}\xi-\chi\xi^{'}-{1\over2}\chi \partial_x^2(-\partial^2+M^2)^{-1/2}\chi\bigr)}.
\ee

Hence
\begin{eqnarray}
\label{bos2}
-{1\over\sqrt\pi}\xi^{'}\Psi[\xi]&=&{1\over\sqrt\pi}{\delta\over\delta\chi}\Psi[\xi,\chi]|_{\chi=0}\nonumber\\
&=&\int Du^{\dagger }Due^{\int d^2xu^{\dagger }\gamma^5u}\hat\psi^\dagger\hat\psi\Psi [u,{\sqrt\pi\over2}\xi u^{\dagger },0].
\end{eqnarray}
If we represent \(\dot\xi\) as \(-i{\partial\over\partial\xi}\) then
(\ref{bos}) and (\ref{bos2}) imply that 
\begin{eqnarray}
&&\int Du^{\dagger }Due^{\int d^2xu^{\dagger }\gamma^5u}{\hat{\bar\psi}}\gamma^\mu\hat\psi\Psi [u,{\sqrt\pi\over2}\xi u^{\dagger },0]\nonumber\\&&\qquad\qquad
=\int Du^{\dagger }Due^{\int d^2xu^{\dagger }\gamma^5u}{-1\over\sqrt\pi}\epsilon^{\mu\nu}\partial_\nu\xi\Psi [u,{\sqrt\pi\over2}\xi u^{\dagger },0].
\end{eqnarray}
It follows that 
${\hat{\bar\psi}}\gamma^\mu\hat\psi\sim{-1\over\sqrt\pi}\epsilon^{\mu\nu}\partial_\nu\xi$
on all physical states. This equivalence may be exploited in a number of
ways; for example it immediately allows us to identify the creation and
annihilation operators of the theory
\be
a_\pm(p)={1\over\sqrt2}\left(\omega^{1/2}j_0(p)/p\mp\omega^{-1/2}j_5(p)\right),
\ee
where $j_0(p)$ and $j_5(p)$ are the Fourier transforms of
$j_0(x)=\hat\psi^\dagger(x)\hat\psi(x)$ and
$j_5(x)=\hat\psi^\dagger(x)\gamma^5\hat\psi(x)$ respectively. 
Also, in calculating physical expectation values 
\be
\langle\Psi_1|O(\hat\psi,\hat\psi^\dagger)|\Psi_2\rangle
\ee
the operator $O$ can be represented in terms of equivalent bosonic operators;
the resulting calculations are generally much easier. To illustrate this,
consider the chiral condensate
\begin{eqnarray}
\label{cc}
\langle\hat\psi^\dagger\hat\psi\rangle_\theta &=&
  {\langle\theta|{\hat{\bar{\psi}}}\hat\psi|\theta\rangle\over\langle\theta|\theta\rangle}\nonumber\\ &=& {\langle0|{\hat{\bar{\psi}}}
  e^{i\theta\gamma^5}\hat\psi|0\rangle\over\langle0|0\rangle},
\end{eqnarray}
where we have used (\ref{vacrep}). Now the bosonic operators corresponding to
the chiral densities $\hat{\bar{\psi}}(1\pm\gamma^5)\hat\psi$ are
\be
{Me^\gamma\over2\pi}:e^{\pm i\sqrt{4\pi}\xi}:
\ee
(this can be verified in a similar way to the other bosonization formulae). The normal-ordering is perfomed with respect to the scale $M$, and $\gamma$
is the Euler constant. Thus $\hat{\bar{\psi}}
  e^{i\theta\gamma^5}\hat\psi$ corresponds to
  ${Me^\gamma\over2\pi}:\cos(\sqrt{4\pi}\xi+\theta):$ whose vacuum
  expectation value is easily ascertained from (\ref{boson}). We have
\be
\langle\hat\psi^\dagger\hat\psi\rangle_\theta={Me^\gamma\over2\pi}\cos\theta,
\ee
which is the well-known result \cite{12}.

\section{The Local Expansion}

\subsection{The Schr\"odinger Functional}

We will now describe how the result (\ref{vac}) can be reconstructed from a derivative
expansion, whose terms are local expressions in the fields. This allows the
techniques of \cite{4,4.5}, which are generalizable to higher dimensions, to
be used to provide an alternative solution of the theory. Naively,
the presence of a mass gap in the theory should mean that all propagators are
exponentially damped at large distances, so that the logarithm of (\ref{vac})
should reduce for slowly varying fields to a sum
of local terms, ie. integrals over finite powers of the fields and their
derivatives, evaluated at a single spatial point. Unfortunately, because of gauge-invariance (\ref{vac}) contains 
massless modes not appearing in the physical spectrum, and this
simplification of the vacuum WF does not occur. This may be seen by noting
the existence of screened large-distance configurations.

We will get around this problem by considering instead the Schr\"odinger
functional 
\begin{equation}
\label{schro}
\Psi _{\tau }[\tilde{u},\tilde{u}^{\dagger },u,u^{\dagger }]=\langle \tilde{u},\tilde{u}^{\dagger }|e^{-H\tau }|u,u^{\dagger }\rangle .
\end{equation}
The vacuum functional is just the \( \tau \rightarrow \infty  \) limit of this
object. We can also extract excited states by inserting a basis of energy
eigenstates 
\begin{equation}
\label{states}
\Psi _{\tau }[0,0,u,u^{\dagger }]\sim \sum _{E}\Psi _{E}[u,u^{\dagger }]e^{-E\tau }.
\end{equation}
The parameter \( \tau  \), which corresponds to Euclidean time, acts as an
inverse mass for all states of the theory---physical or otherwise. Provided
we work with fields that vary slowly on the scale of $\tau$, the logarithm of
$\Psi_\tau$ has an
expansion in positive powers of \( \tau  \), each term of which is itself a
finite sum of local terms. As we
will see, the
large \( \tau  \) behaviour can be reconstructed from this expansion. 

We will begin by finding the free fermion solution. The Schr\"odinger equation
is
\begin{equation}
\label{tschr}
-\frac{\partial }{\partial \tau }\Psi _{\tau }=\hat{H}\Psi _{\tau },
\end{equation}
where
\begin{equation}
\label{hamil}
\hat{H}=\frac{1}{2}\int d^{d}x(u^{\dagger }+\frac{\delta }{\delta u})h(u+\frac{\delta }{\delta u^{\dagger }}),\qquad h=i\gamma ^{5}\partial _{x}.
\end{equation}
We also have the initial condition 
\begin{equation}
\label{initial}
\Psi _{0}=\langle Q|u,u^{\dagger }\rangle\!\rangle =e^{\int dx(u^{\dagger }Qu)},
\end{equation}
If we put 
\begin{equation}
\label{gamma}
\Psi _{\tau }=e^{\int dx(u^{\dagger }\Gamma (\tau )u)},
\end{equation}
then (\ref{sch}) becomes
\begin{equation}
\label{geqn}
\dot{\Gamma }=-\frac{1}{2}(1-\Gamma )h(1+\Gamma ),
\end{equation}
with \( \Gamma (0)=Q \). This equation is solved by
\begin{equation}
\label{ss}
\Gamma =(\Sigma +Q_{-})(\Sigma -Q_{-})^{-1},
\end{equation}
and substitution into (\ref{geqn}) gives \( \dot{\Sigma }=h\Sigma  \), \( \Sigma (0)=Q_{+} \),
with the solution
\begin{equation}
\label{sols}
\Sigma =e^{-i\gamma ^{5}\tau \partial _{x}}Q_{+}.
\end{equation}
Hence we obtain
\begin{equation}
\label{ups}
\Gamma =Q+2Q_{-}\gamma ^{5}\tanh (-i\tau \partial _{x})Q_{+},
\end{equation}
 which has a derivative expansion for small \( \tau  \), as promised. It is
also easy to find the (non-local) large-time limit:
\begin{equation}
\label{limit}
\lim _{\tau \rightarrow \infty }\Gamma (x-y)=Q+2Q_{-}\frac{i\gamma ^{5}}{\pi }{\cal P}\frac{1}{x-y}Q_{+},
\end{equation}
 which coincides with the solution \( \Gamma =A^{-1}C \) that we found before.

Now consider the interacting theory. Substituting the result (\ref{sch}) into the time-dependent
Schr\"odinger equation (\ref{tschr}) gives
\begin{eqnarray}
-\dot{f}^{a}(x_{1},y_{1},\ldots ,x_{a},y_{a}) & = & -E_{0}f^{a}(x_{1},y_{1},\ldots ,x_{a},y_{a})+V(x_{1},y_{1},\ldots ,x_{a},y_{a})f^{a}(x_{1},y_{1},\ldots ,x_{a},y_{a})\nonumber \\
 &  & -i\int dx\lim _{y\rightarrow x}\partial _{y}f^{a+1}(x_{1},y_{1},\ldots ,x_{a},y_{a},y,x)\nonumber \\
 &  & +i\int dx\lim _{y\rightarrow x}\partial _{y}\sum _{b=1}^{a}f^{a+1}(x_{1},y_{1},\ldots ,x_{b},x,y,y_{b},\ldots ,x_{a},y_{a})\nonumber \\
 &  & -i\sum _{b=1}^{a}f^{a-1}(x_{1},y_{1},\ldots ,\rlap /x_{b},\rlap /y_{b},\ldots ,x_{a},y_{a})\delta ^{\prime }(x_{b}-y_{b}),\label{tsch} 
\end{eqnarray}
Here we have made \( \hat{H} \) regular by subtracting the zero-point energy
(\ref{energy}), ie. \( \hat{H}\rightarrow \hat{H}-E_{0} \); this may be
achieved by normal-ordering. In order that we recover the
solution (\ref{ups}) in the free-fermion limit \( e\rightarrow 0 \), we use
the Ansatz (\ref{ans}) with \( \Delta (x-y)=\tanh (i\tau \partial _{x})\delta (x-y) \).
The \( a=0 \) part of (\ref{tsch}) is of course trivial; for \( a=1 \) it
becomes
\begin{eqnarray}
0 & = & \Delta (x_{1}-y_{1})\partial _{\tau }\Phi (x_{1}-y_{1})+V(x_{1}-y_{1})\Delta (x_{1}-y_{1})\nonumber \\
 &  & +i\int dx\Delta (x_{1}-x)\Delta (x-y_{1})(\Phi ^{\prime }(x-y_{1})-\Phi ^{\prime }(x-x_{1}))\nonumber \\
 &  & -i\Delta (x_{1}-y_{1})\int dx\lim _{y\rightarrow x}\partial _{y}\left( \Delta (y-x)e^{g(x,y,x_{1},y_{1})}\right) -E_{0}\Delta (x_{1}-y_{1}).\label{any} 
\end{eqnarray}
In obtaining this we used the identity
\begin{eqnarray}
\dot{\Delta }(x_{1}-y_{1}) & = & i(1-\tanh ^{2}(i\tau \partial ))\delta ^{\prime }(x_{1}-y_{1})\nonumber \\
 & = & i\delta ^{\prime }(x_{1}-y_{1})-i\int dx\Delta (x_{1}-x)\Delta^\prime (x-y_{1}).
\end{eqnarray}
Now for small \( x \)
\begin{equation}
\label{tanh}
\Delta (x)=\frac{i\pi }{\tau }{\mathrm{cosech}}\left( \frac{\pi ^{2}x}{\tau }\right) \sim \frac{i}{\pi x}+O(x),
\end{equation}
so that 
\begin{eqnarray}
\lim _{x\rightarrow 0}\Delta (x)f(x)&=&\frac{i}{\pi }f^{\prime }(0)+\ldots\nonumber\\
\lim _{x\rightarrow 0}\Delta ^{\prime }(x)f(x)&=&\frac{-i}{2\pi }f^{\prime \prime }(0)+\ldots
\end{eqnarray}
The ellipses have the same meaning as in (\ref{lim}). Hence as before
\be
-i\int dx\lim _{y\rightarrow x}\partial _{y}\left( \Delta (y-x)e^{g(x,y,x_{1},y_{1})}\right)  =  E_{0}-\int dpp^{2}C(p,\tau)^{2}(1-\cos(p(x_1-y_1))). 
\ee
On the other hand,
we find that
\begin{eqnarray}
i\int dx\Delta (x_{1}-x)\Delta (x-y_{1}) & (\Phi ^{\prime }(x-y_{1})-\Phi ^{\prime }(x-x_{1}))= & \nonumber \\
 & 2\Delta (x_{1}-y_{1})\int dp\coth (\tau p)pC & (p,\tau)(1-\cos (p(x_{1}-y_{1}))).\label{newiden} 
\end{eqnarray}
Thus (\ref{any}) reduces to
\begin{equation}
\label{diff}
\dot{C}(p,\tau )+2p\coth (p\tau )C(p,\tau )-p^{2}C^{2}(p,\tau)+\frac{M^{2}}{p^{2}}=0.
\end{equation}
The initial condition (\ref{initial}) is satisfied if \( C(p,0)=0 \). For
small $\tau$ we can
expand \( C \) as
\begin{equation}
\label{cseries}
C=\sum _{n =1}^{\infty }c_{n}\tau ^{n},
\end{equation}
and substituting into (\ref{diff}) leads to a recursion relation which is
easily solved for the coefficients $c_n$, which are polynomials in positive powers
of \( p \), divided by  \( 1/p^{2} \). We note
that (\ref{diff}) reduces to (\ref{quad}) for large \( \tau  \), so that we recover
the vacuum functional, as expected.

Just as for the vacuum functional, it is easily shown that the solution given by (\ref{diff}) satisfies (\ref{tsch}) for all \( a \).

\subsection{Solution from a local Ansatz}

The Ansatz (\ref{ans}) is very useful in two dimensions, but it is not clear
that it generalizes in any way to higher dimensions. Thus we will now describe
an Ansatz which {\em does\/} generalize.

Provided we work with fields that vary slowly on the scale of \( \tau  \),
\( \Delta (x)=\tanh (i\partial _{x}\tau )\delta (x) \) can be expanded in
derivatives 
of the delta function. It follows from (\ref{ans}) that the logarithm of \(
\Psi _{\tau } \) has a local expansion (this should also be true in higher
dimensions). This allows us to 
 write \( \log \Psi _{\tau } \) in the form (\ref{wf}), where we take the \( f^{a} \)
to be local:

\begin{eqnarray}
f^{a}(p_{1},\ldots ,p_{2a}) & = & \sum _{n_{1},\ldots ,n_{2a}=0}^{\infty }b_{n_{1}}p_{1}^{n_{1}}\ldots b_{n_{2a}}p_{2a}^{n_{2a}}\nonumber \\
f^{a}(x_{1},y_{1},\ldots ,x_{a},y_{a}) & = & \int \frac{dp_{1}}{2\pi }\ldots \frac{dp_{2a}}{2\pi }\delta (p_{1}+\ldots +p_{2a})f^{a}(p_{1},\ldots ,p_{2a}).\label{fmom} 
\end{eqnarray}
Since we can gauge away the electromagnetic field, this leads to the
following Ansatz

\begin{eqnarray}
\label{lanz}
\Psi _{\tau }&=&1+\int dx\sum _{n,q=0}^{\infty }a_{n}^{(q)}\tau^qu^{\dagger
  }(x)\gamma ^{5}\partial _{x}^{n}u(x)\nonumber\\
& &+\int dxdy\sum _{n,m,q=0}^{\infty }a^{(q)}_{nm}\tau^qu^{\dagger }(x)\gamma ^{5}\bigl(\partial _{x}^{n}u(x)\bigr)u^{\dagger }(y)\gamma ^{5}\bigl(\partial _{y}^{m}u(y)\bigr)+\ldots, 
\end{eqnarray}
where the coefficients $a_{ij\ldots}$ are obtained from the $b_n$ in the obvious
way. This Ansatz can be inserted directly into the Schr\"odinger
equation (\ref{tsch}), which thus reduces to an infinite set of algebraic
equations (a finite set for each order in $\tau$). We have explicitly
verified for the first few terms the that the local
expansion which results from solving the Schr\"odinger equation in this way
is the same as that which is obtained by expanding the small-$\tau$ solution
found in the last section.

Now the local expansion depended on expanding $\Psi_\tau$ in positive powers
of $\tau$ for small $\tau$, but we can reconstruct the large $\tau$ behaviour
from a knowledge of the local expansion alone. If we evaluate $\Psi_\tau$ for
scaled fields $u(x/\sqrt\rho)$ and $u^\dagger(x/\sqrt\rho)$ then it can be
analytically continued to the complex $\rho$-plane with the negative real
axis removed. The proof of this is a straightforward generalization of the
arguments of \cite{4,4.5}. Then by using
Cauchy's theorem, we can relate the value of the 
functional at $\rho=1$ to the value at large $\rho$, where the fields are
slowly varying, so that the $\tau$-expansion converges for large
$\tau$. Specifically, we study the following integral:

\be
\label{lsch}
{1\over2\pi i}\int_C {d\rho\over\rho-1}e^{\lambda\rho}\Psi_\tau[u(x/\sqrt\rho,u^\dagger(x/\sqrt\rho)].
\ee
If we take $C$ to be a circle of sufficiently large radius we can evaluate this integral by
inserting the local expansion of $\Psi_\tau$ for any value of $\tau$. Alternatively, we can collapse
the contour around the point $\rho=1$, and the negative real axis, the latter
contribution being exponentially suppressed for large $\lambda$. Thus we have
successfully expressed $\Psi_\tau$ for large $\tau$ in terms of the local
expansion. In practise we will wish to
truncate the expansion at some order, in which case the truncation
error is minimized by taking $\lambda$ to be large but finite \cite{4}.

The Ansatz (\ref{fmom}) generalizes straightforwardly to higher dimensions,
and could form the starting point for a numerical approach to (3+1)
dimensional QCD.

\section{Discussion}

In this paper we have shown how the degeneracy of the Floreanini-Jackiw
representation for fermions is related to a choice of diagonalization of the
fermionic fields. We have used this to derive formal expressions for
gauge-invariant 
wave-functionals in the Schr\"odinger representation, and obtained the vacuum
wave-functional of the Schwinger model, both from a path-integral
expression and as a solution of the Schr\"odinger equation. All known
features of the model arise naturally in this context.

In general, the wave-functionals of a theory with a mass gap should be
expandable in local quantities at large distances. We have seen how the
existence of screened large-distance configurations spoils this feature in
the Schwinger model. 
However, we have obtained the Schr\"odinger wave-functional and shown that
it does have an expansion in local quantities. This may be
obtained directly from the Schr\"odinger equation, which
reduces to a set of algebraic equations. An analogous technique can be applied to gauge theories in higher dimensions, where the Schr\"odinger
functional should also have such an expansion. Physical wave-functionals may be
reconstructed by considering its behaviour for large values of the time
parameter, which acts as an infrared regulator.

The techniques of this paper generalize in a straightforward way to other two dimensional theories, for example involving massive fermions or a non-abelian gauge group. Various perturbative and non-perturbative approximation schemes present themselves. And though the detailed structure of the solutions is specific to two dimensions, the solution via a local expansion is a technique which should work in a similar way for higher dimensional gauge theories.

The main steps involved in implementing a numerical approach to higher
dimensional $QCD$ based on solving the Schr\"odinger equation for the local
expansion are as follows. First, to obtain the generalization of (\ref{gauss})
to non-Abelian fields; this is straightforward, and allows the Hamiltonian to
be written in terms of gauge-invariant currents. Second, to write down the local ansatz which generalizes (\ref{lanz}). The local gauge-invariant terms which make up the local expansion are easily identified. Renormalization may be dealt with by adding counterterms familiar from perturbation theory. Or, we can
consider supersymmetric theories in which they are absent. Finally, after finding the action of the Hamiltonian on the terms of the local expansion, the
solution of the Schr\"odinger equation proceeds as described in section 4
above.  

\section*{Acknowledgements}
The author acknowledges a studentship from EPSRC.

\end{document}